\documentclass[10pt,aps,prd,preprintnumbers,showpacs,nofootinbib,superscriptaddress,notitlepage,twocolumn]{revtex4-1}

\usepackage{graphicx}
\usepackage{color}
\usepackage{amsmath}
\usepackage{amssymb}
\usepackage{enumerate}
\usepackage[colorlinks=true,citecolor=blue,urlcolor=blue]{hyperref}

\newcommand{\PRE}[1]{{#1}} 
\newcommand{\tot}{\textrm{tot}}
\newcommand{\halo}{\textrm{halo}}
\newcommand{\MW}{\textrm{MW}}
\renewcommand{\max}{\textrm{max}}

\begin{document}

\title{Angular distribution of gamma-ray emission from velocity-dependent dark matter annihilation in subhalos}

\author{Kimberly K.~Boddy}
\affiliation{\mbox{Department of Physics \& Astronomy,
Johns Hopkins University, Baltimore, MD 21218, USA}}

\author{Jason Kumar}
\affiliation{\mbox{Department of Physics \& Astronomy,
University of Hawai`i, Honolulu, HI 96822, USA}}

\author{Jack Runburg}
\affiliation{\mbox{Department of Physics \& Astronomy,
University of Hawai`i, Honolulu, HI 96822, USA}}

\author{Louis E.~Strigari\PRE{\vspace*{.1in}}}
\affiliation{{Department of Physics and Astronomy,
Mitchell Institute for Fundamental Physics and Astronomy,
Texas A\&M University, College Station, TX  77843, USA}}

\begin{abstract}
\PRE{\vspace*{.1in}}

We consider the effect of velocity-dependent dark matter annihilation on the angular distribution of gamma rays produced in dark matter subhalos. We assume that the dark matter potential is spherically symmetric, characterized by a scale radius and scale density, and the velocity distribution is isotropic. We find that the effect of velocity-dependent dark matter annihilation is largely determined by dimensional analysis; the angular size of gamma-ray emission from an individual subhalo is rescaled by a factor which depends on the form of the dark matter distribution, but not on the halo parameters, while the relative normalization of the gamma-ray flux from different mass subhalos is rescaled by a factor which depends on the halo parameters, but not on the form of the dark matter distribution. We apply our results to a Navarro-Frenk-White profile for the case of an individual subhalo and comment on the application of these results to a distribution of subhalos.

\end{abstract}

\maketitle

\section{Introduction}

Gamma-ray experiments~\cite{Conrad:2015bsa} and the cosmic microwave background~\cite{Aghanim:2018eyx} place strong constraints on dark matter annihilation into visible particles.
Standard cold dark matter particles move nonrelativistically both in the early Universe and within collapsed structures; thus, these observations are particularly sensitive to the $s$-wave component of the annihilation cross section, which is independent of the dark matter relative velocity.
Fermi-LAT studies~\cite{Ackermann:2015zua,Geringer-Sameth:2014qqa} and ground-based observations~\cite{Ahnen:2016qkx,Albert:2017vtb} of dwarf spheroidal galaxies (dSphs) place the strongest gamma-ray bounds on this cross section.
In addition to dSphs, gamma-ray signals from dark matter annihilation may also arise in other interesting astrophysical systems, including the Galactic center and M31~\cite{Karwin:2019jpy}, dark matter subhalos~\cite{Ackermann:2012nb}, galaxy clusters~\cite{Ackermann:2015fdi}, and the diffuse Galactic and extragalactic background~\cite{Ajello:2015mfa}.

There has been recent renewed interest in extending these limits by examining the effects of velocity-dependent dark matter annihilation on the flux of gamma rays emitted by astrophysical objects with high dark matter densities~\cite{Robertson:2009bh,Ferrer:2013cla,Boddy:2017vpe,Zhao:2017dln,Petac:2018gue,Boddy:2018ike,Lacroix:2018qqh,Belotsky:2014}.
In particular, for velocity-dependent annihilation, the standard factorization of the gamma-ray flux into an astrophysical contribution (the $J$-factor) and a particle physics contribution fails, because the dark matter annihilation cross section is nontrivially correlated with the dark matter velocity distribution.
The velocity dependence of the annihilation can be absorbed into the calculation of the $J$-factor, impacting the total $J$-factor of dSphs, as well as the angular distribution of the gamma-ray flux from annihilation in the Galactic Center.

In this paper we extend upon these previous analyses and study the effect of velocity-dependent dark matter annihilation on the angular distribution of gamma-ray emission from small subhalos.
We find that a key simplification occurs if the only dimensionful constants which enter the expression for the dark matter velocity distribution are a scale radius, a scale density, and Newton's constant.
The effect of velocity-dependent annihilation is then largely determined by dimensional analysis.
We consider the cases of $s$-, $p$-, and $d$-wave annihilation, as well as Sommerfeld-enhanced annihilation in the Coulomb limit; in all of these cases, the dark matter annihilation cross section has a power-law dependence on the relative velocity.
We can define a scale free angular distribution of the gamma-ray flux such that a change to the power-law behavior alters it from any particular subhalo in a manner that is independent of the halo parameters or the distance to the halo, although the normalization of the flux does depend on the profile parameters.

We apply these results to the case of a Navarro-Frenk-White (NFW) profile and consider the effects of velocity-dependent annihilation on the angular distribution of the resulting gamma-ray emission from a single subhalo.
We then comment on the effect of velocity dependence in a population of unresolved subhalos.
Dark matter signals from a subhalo population are especially interesting, because they can lead to anisotropies in the gamma-ray spectrum observed by instruments such as Fermi-LAT~\cite{Fornasa:2016ohl,Ackermann:2018wlo} and CTA~\cite{Hutten:2018wop}.
Moreover, numerical simulations and gravitational lensing studies are expected to provide increasingly sophisticated models for the probability distribution function of subhalos~\cite{Hezaveh:2016ltk,Cyr-Racine:2018htu}.
With this distribution function, it would be possible to estimate the expected angular power spectrum of the gamma-ray flux.
However, we note that such an estimate would depend crucially on the assumptions made about the velocity dependence of dark matter annihilation: velocity-dependent annihilation affects the angular size of the gamma-ray emission from individual subhalos, as well as the relative intensity of the flux from different subhalos.

The plan of this paper is as follows.
In Sec.~\ref{sec:Jeff}, we describe the formalism for determining the angular distribution of the gamma-ray flux arising from velocity-dependent dark matter annihilation in a small subhalo, and we consider specific models of annihilation in Sec.~\ref{sec:ann}.
In Sec.~\ref{sec:subhalo-single}, we apply this formalism to the particular case of a single halo with an NFW profile.
In Sec.~\ref{sec:subahlo-distribution}, we comment on the impact of these results for a population of subhalos.
We conclude in Sec.~\ref{sec:conclusion}.

\section{Effective \texorpdfstring{$J$}{J}-factor of a subhalo}
\label{sec:Jeff}

We consider dark matter annihilation with a cross section of the form $\sigma v = (\sigma v)_0 \times S(v_r/c)$, where $(\sigma v)_0$ is a dimensionful constant and $v_r$ is the relative velocity of the incoming particles.
In the simplest scenario, $S(v_r/c)=1$ and the resulting differential gamma-ray flux takes the standard form $d\Phi/d\Omega = \Phi_\textrm{PP} J(\Omega)$, where $\Phi_\textrm{PP} \propto (\sigma_A v)_0$ depends only on the particle physics properties of dark matter.
Additionally, the $J$-factor relies only on the astrophysical properties of the dark matter halo and has the form $\int d\ell ~\rho^2(\boldsymbol{r})$, where $\ell$ is the distance along the line of sight and $\rho(\boldsymbol{r})$ is the dark matter density profile of the halo at the location $\boldsymbol{r}$ from the halo center.

For the more general case of $S(v_r/c) \neq 1$, factoring the particle physics and astrophysics contributions to the gamma-ray flux is no longer possible.
We thus incorporate $S(v_r/c)$ into the calculation of the $J$-factor by writing the density profile as an integral over the velocity distribution of particles in the halo: $\rho(\boldsymbol{r}) = \int d^3v ~f(\boldsymbol{r},\boldsymbol{v})$.
This effective $J$-factor at the location $(\theta,\phi)$ from the halo center on the sky is
\begin{align}
  J_S (\theta,\phi) = \int_0^\infty & d\ell \int d^3 v_1 \int d^3 v_2 \
  S(|\boldsymbol{v}_1-\boldsymbol{v}_2|/c) \\\nonumber
  &\times
  f\left[\boldsymbol{r}(\ell, \theta),\boldsymbol{v}_1\right] \
  f\left[\boldsymbol{r}(\ell, \theta),\boldsymbol{v}_2\right].
\end{align}
Note that $r^2(\ell, \theta) = \ell^2 + D^2 -2\ell D \cos \theta$, where $D$ is the distance to the center of the halo.
The total gamma-ray flux $\Phi$ is proportional to the integrated $J$-factor, given by $J_S^\tot = \int d\Omega ~J_S (\theta,\phi)$.

The calculation of the effective $J$-factor simplifies greatly by assuming the dark matter distribution is spherically symmetric (in which case the $J$-factor is a function of $\theta$ only) and has isotropic orbits.
The dark matter density and velocity distributions then depend only on the distance from the halo center $r$ and the magnitude of the relative particle velocities $v$.
We thus define the dimensionless radius, density profile, and velocity as
\begin{equation}
  \tilde{r} \equiv \frac{r}{r_s},\
  \tilde{\rho}(\tilde{r}) \equiv \frac{\rho(r)}{\rho_s},\ \textrm{and}\
  \tilde{v} \equiv \frac{v}{\sqrt{4\pi G_N \rho_s r_s^2}}
  \label{eq:dimensionless-scales}
\end{equation}
to write the scale free velocity distribution $\tilde f$ as
\begin{equation}
  \tilde{f}(\tilde{r},\tilde{v}) = (4\pi G_N)^{3/2} r_s^3 \rho_s^{1/2} f(r,v),
  \label{eq:distribution-scale}
\end{equation}
where $\tilde{\rho} (\tilde{r}) = \int d^3\tilde{v} ~\tilde{f}(\tilde{r},\tilde{v})$.

Since we are interested in the $J$-factor for small subhalos, we make the additional assumption that the characteristic size of any given subhalo is much smaller than its distance away: $r_s \ll D$.
As long as the dark matter annihilation rate falls off rapidly for $r>r_s$ (which is the case we consider in Sec.~\ref{sec:subhalo-single}), most of the associated gamma-ray emission occurs close the halo center at $\theta \ll 1$.
We define the quantity $\tilde{\theta} \equiv \theta / \theta_0$, where $\theta_0 \equiv r_s / D$.
We may express the total effective $J$-factor as
\begin{equation}
  J_S^\tot \approx 2\pi\theta_0^2 \int_0^\infty
  d\tilde{\theta}\ \tilde{\theta}J_S(\tilde{\theta}),
\end{equation}
where we extend the upper limit of integration to infinity since $\theta_0 \ll 1$, though most of the contribution to the integral occurs for $\tilde{\theta} \lesssim 1$.

\section{Annihilation Models}
\label{sec:ann}

We now build upon the formalism for the effective $J$-factor by specifying the velocity dependence of the annihilation cross section.
We consider a class of annihilation models with the form $S(v_r/c) = (v_r/c)^n$, where $n$ is an integer.
In particular, we are interested in the following possibilities:
\begin{itemize}
\item $n=0$: Dark matter annihilates from an $s$-wave initial state.
\item $n=2$: Dark matter is a Majorana fermion that annihilates to Standard
Model fermion/anti-fermion pairs through an interaction respecting minimal flavor violation (e.g. Ref.~\cite{Kumar:2013iva}).
  Annihilation from an $s$-wave initial state is chirality-suppressed, and dark
matter instead annihilates from a $p$-wave initial state.
\item $n=4$: This scenario is similar to that for $n=2$, except dark matter is a real scalar particle~\cite{Giacchino:2013bta,Toma:2013}.
  In this case, the $p$-wave initial state is forbidden, and dark matter instead annihilates from a $d$-wave initial state~\cite{Kumar:2013iva,Giacchino:2013bta,Toma:2013}.
\item $n=-1$: Dark matter self-interacts through a relatively long-ranged force.
  The annihilation cross section is Sommerfeld-enhanced~\cite{ArkaniHamed:2008qn,Feng:2010zp}, and $\sigma v \propto 1/v$ in the Coulomb limit.
\end{itemize}

We may now write the effective $J$-factor and the total effective $J$-factor for a particular value of $n$ as
\begin{align}
  J_{S(n)}(\tilde{\theta}) &= 2r_s \rho_s^2
  \left(\frac{4\pi G_N \rho_s r_s^2}{c^2}\right)^{n/2}
  \tilde{J}_{S(n)} (\tilde{\theta}) \label{eq:J} \\
  J_{S(n)}^\tot &= \frac{4\pi r_s^3 \rho_s^2}{D^2}
  \left(\frac{4\pi G_N \rho_s r_s^2}{c^2}\right)^{n/2}
  \tilde{J}_{S(n)}^\tot , \label{eq:Jtot}
\end{align}
where the corresponding dimensionless quantities are
\begin{align}
  \tilde{J}_{S(n)} (\tilde{\theta}) &\approx \int_{\tilde{\theta}}^\infty d\tilde{r}
  \left[1-\left(\frac{\tilde{\theta}}{\tilde{r}}\right)^2\right]^{-1/2}
  P^2_n(\tilde{r}) \label{eq:Jtilde} \\
  \tilde{J}_{S(n)}^\tot &\approx \int_0^\infty
  d\tilde{\theta}\ \tilde{\theta}\tilde{J}_{S(n)}(\tilde{\theta})
  \label{eq:Jtottilde}
\end{align}
in the $r_s \ll D$ limit.
The quantity $P^2_n(\tilde{r})$ is analogous to $\tilde{\rho}^2(r)$ in the standard $J$-factor calculation and is given by
\begin{align}
  P^2_n(\tilde{r}) \equiv& \int d^3 \tilde{v}_1\ d^3 \tilde{v}_2\
  |\tilde{\boldsymbol{v}}_1 - \tilde{\boldsymbol{v}}_2|^n
  \tilde{f}(\tilde{r},\tilde{v}_1) \tilde{f}(\tilde{r},\tilde{v}_2)
  \nonumber\\
  =& 8\pi^2 \int_0^\infty d\tilde{v}_1 \int_0^\infty d\tilde{v}_2
   \int_{|\tilde{v}_1-\tilde{v}_2|}^{\tilde{v}_1+\tilde{v}_2} d\tilde{v}_r \nonumber\\
  &\times \tilde{f}(\tilde{r},\tilde{v}_1)\ \tilde{f}(\tilde{r},\tilde{v}_2)\
  \tilde{v}_1\ \tilde{v}_2\ \tilde{v}_r^{n+1},
  \label{eq:P2}
\end{align}
with $P^2_n$ reducing to $\tilde{\rho}^2$ for $n=0$.
We note that the overall scaling dependence of the halo parameters in Eq.~\eqref{eq:Jtot} is consistent with that found in Ref.~\cite{Boddy:2017vpe}, which considered Sommerfeld-enhanced annihilation.

The strategy for discriminating between annihilation models in gamma-ray data is different if the analysis is focused on a single subhalo or if it incorporates a population of subhalos.
In Sec.~\ref{sec:subhalo-single}, we find that for an individual halo, it is necessary to study the angular dependence of the gamma-ray emission.
However, for many subhalos, the relative amplitude of the gamma-ray flux between subhalos can be used to distinguish annihilation models, as we discuss in Sec.~\ref{sec:subahlo-distribution}.

\section{Gamma-ray emission from a single subhalo}
\label{sec:subhalo-single}

We first focus on an individual subhalo for different dark matter annihilation models.
The total gamma-ray flux, which is proportional to the total effective $J$-factor in Eq.~\eqref{eq:Jtot}, from annihilation in the subhalo provides no information on the velocity dependence of the annihilation: changing $n$ results in an overall scaling of the total effective $J$-factor that can be compensated for by adjusting $(\sigma v)_0$, which controls the normalization of the gamma-ray flux.

The velocity dependence can instead be determined from the scale free angular distribution, $\tilde{J}_{S(n)} (\tilde{\theta}) / \tilde{J}_{S(n)}^\tot$, which depends only on $n$ and the scale free velocity distribution $\tilde{f}(\tilde{r},\tilde{v})$.
Since the subhalo is assumed to have a small angular size, a gamma-ray instrument may not have the angular resolution to map out the full angular distribution at high precision.
However, we assume there is enough resolution to identify the average angular size of the gamma-ray emission, which is approximately
\begin{equation}
  \langle\theta\rangle = \theta_0 \frac{\int_0^\infty d\tilde{\theta}\
  \tilde{\theta}^2 \tilde{J}_{S(n)}(\tilde{\theta})}{\tilde{J}_{S(n)}^\tot} .
  \label{eq:angular_size}
\end{equation}
The quantity $\langle\theta\rangle/\theta_0$ is independent of any halo scale parameters and instead depends only on the velocity dependence of the annihilation cross section and on the scale free velocity distribution.

In order to demonstrate the sensitivity of the size of gamma-ray emission on the annihilation model, we must know the form of the scale free velocity distribution.
The remainder of the section describes the procedure for obtaining the velocity distribution using Eddington's formula and then applies the effective $J$-factor formalism to the specific case of a halo with an NFW profile.

\subsection{Dark matter velocity distribution}

We assume a dark matter halo in equilibrium with a spherically symmetric potential.
Jeans's theorem (or equivalently, Liouville's theorem) allows us to write the phase-space distribution function in terms of the integrals of motion~\cite{Widrow:2000dm}.
For dark matter with isotropic orbits, the distribution function depends only on the energy per unit mass $E = v^2/2 + \Phi(r) < 0$, where $\Phi(r) < 0$ is the gravitational potential (assumed to vanish as $r\to\infty$).
We may express the distribution function as $f(E) = f(r,v)$, normalized such that
\begin{align}
  \rho (r) &= 4\pi \int_0^{\sqrt{-2\Phi(r)}} dv~v^2 f(r,v) \nonumber\\
  &= 4\sqrt{2} \pi \int_{\Phi(r)}^0 dE ~ f(E) \sqrt{E - \Phi(r)} ,
\end{align}
where $f(E)$ and $d\rho / d\Phi$ are related by Abel's integral equation, yielding the Eddington inversion formula
\begin{equation}
  f(E) = \frac{1}{\sqrt{8} \pi^2} \int_E^0
  \frac{d^2 \rho}{d\Phi^2} \frac{d\Phi}{\sqrt{\Phi - E}}.
\end{equation}
Using Eq.~\eqref{eq:dimensionless-scales}, we can construct the scale free quantities
\begin{align}
  \tilde{\Phi}(\tilde{r}) &= \frac{\Phi(r)}{4\pi G_N \rho_s r_s^2}
  = -\int_{\tilde{r}}^\infty \frac{dx}{x^2} \int_0^{x} dy~y^2 \tilde{\rho}(y), \nonumber\\
  \tilde{E}(\tilde{r},\tilde{v}) &= \frac{E}{4\pi G_N \rho_s r_s^2}
  = \frac{\tilde{v}^2}{2} + \tilde{\Phi}(\tilde{r})
\end{align}
to obtain
\begin{equation}
  \tilde{f}(\tilde{E}) = \frac{1}{\sqrt{8} \pi^2} \int_{\tilde{E}}^0 \frac{d^2 \tilde{\rho}}{d\tilde{\Phi}^2}
  \frac{d\tilde{\Phi}}{\sqrt{\tilde{\Phi} - \tilde{E}}},
\end{equation}
which is consistent with Eq.~\eqref{eq:distribution-scale}.
Thus, given $\tilde{\rho}(\tilde{r})$, we can perform a numerical integration to obtain $\tilde{f}(\tilde{r},\tilde{v})$.

\begin{figure}[t]
  \centering
  \includegraphics[width=.9\columnwidth]{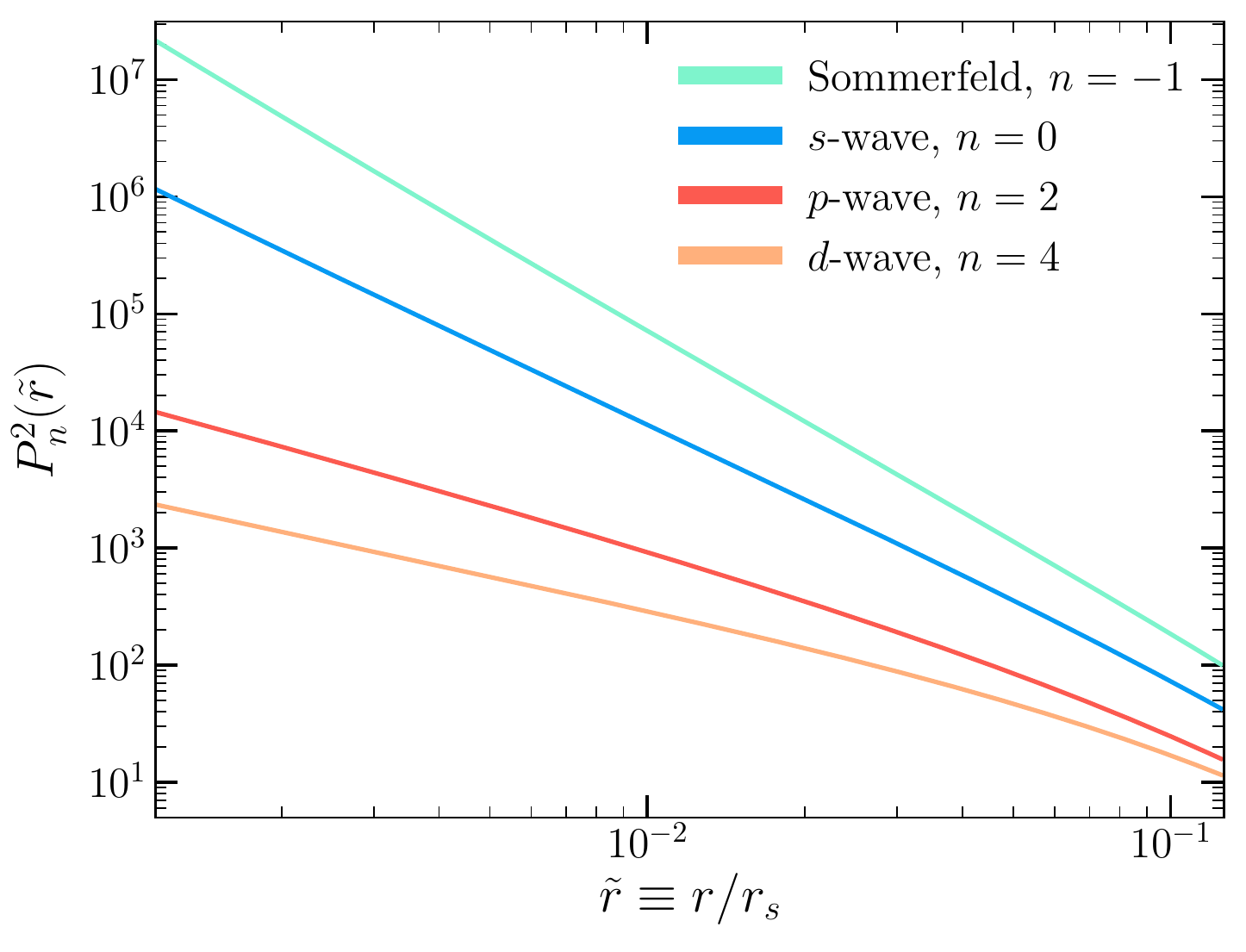}
  \caption{The scale free effective density squared from Eq.~\eqref{eq:P2} for an NFW halo profile.
    It is expressed as a dimensionless quantity in units of $\rho_s^2$ and is a function of the scale free radius $\tilde{r} \equiv r/r_s$.
    We show results for Sommerfeld-enhanced (teal), $s$-wave (blue), $p$-wave (red), and $d$-wave (orange) annihilation.
    For the case of $s$-wave annihilation, this curve is simply $\tilde{\rho}^2(\tilde{r})$.}
  \label{fig:P2}
\end{figure}

\subsection{Application to an NFW profile}

For concreteness, we consider a subhalo with an NFW profile, identifying $r_s$ and $\rho_s$ with the standard NFW scale radius and density parameters to yield $\tilde{\rho}(\tilde{r}) = \tilde{r}^{-1} (1+\tilde{r})^{-2}$.
In Fig.~\ref{fig:P2}, we show the scale free effective halo density squared $P^2_n(\tilde{r})$ for the values of $n=\{-1,0,2,4\}$, corresponding to the annihilation models of interest, described in Sec.~\ref{sec:ann}.

\begin{figure*}[t]
  \includegraphics[scale=0.5]{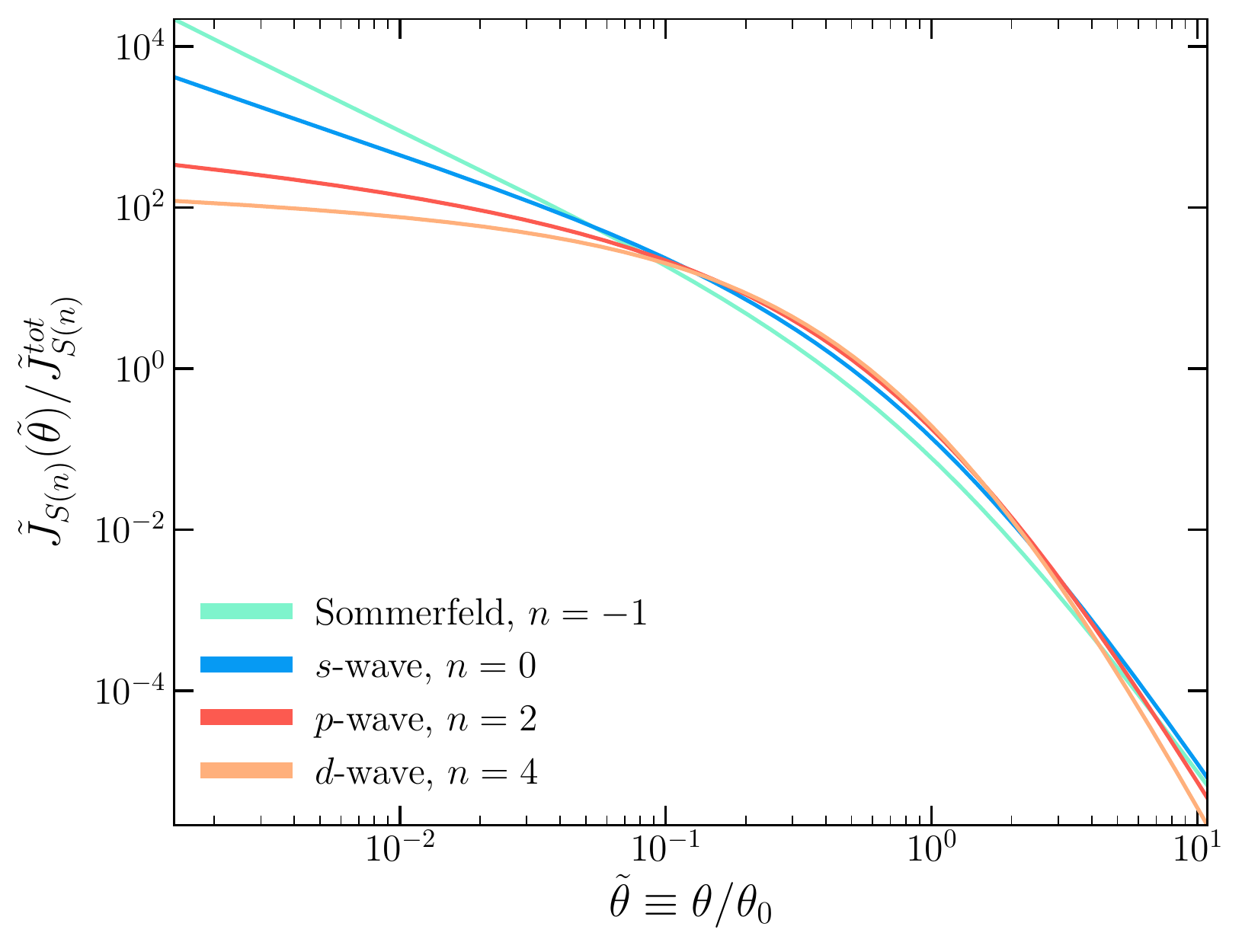}
  \includegraphics[scale=0.5]{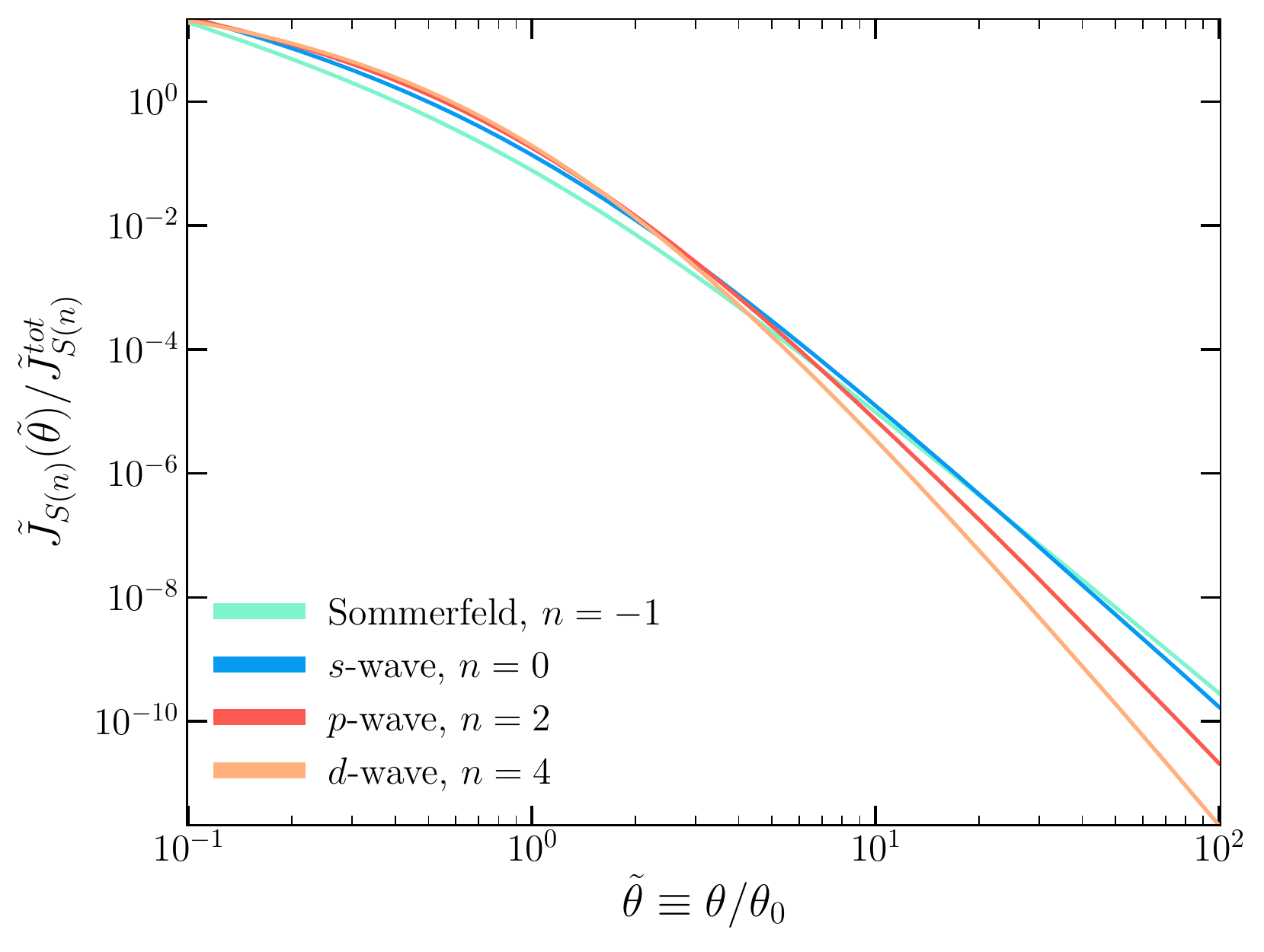}
  \caption{The scale free angular distribution of the gamma-ray flux from dark matter annihilation in an NFW subhalo, as function of $\tilde{\theta} \equiv \theta/\theta_0$, under the approximation $\theta_0 \equiv r_s/D \ll 1$.
    We show separately the distribution at small (left panel) and large (right panel) values of $\tilde{\theta}$.
    As in Fig.~\ref{fig:P2}, the curves represent Sommerfeld-enhanced (teal), $s$-wave (blue), $p$-wave (red), and $d$-wave (orange) annihilation.}
  \label{fig:angular}
\end{figure*}

In Fig.~\ref{fig:angular}, we show the scale free angular distribution, given by $\tilde{J}_{S(n)}(\tilde{\theta})/\tilde{J}_{S(n)}^\tot$, of the gamma-ray flux arising from dark matter annihilation.
The numerical values for $\tilde{J}_{S(n)}^\tot$ are listed in Table~\ref{tab:J_and_theta}.
For the larger values of $n$, the scale free angular distribution is increasingly suppressed at both large and small $\tilde{\theta}$.
This finding is consistent with previous results studying the angular distribution of gamma-ray emission from dark matter annihilation in the Milky Way halo~\cite{Boddy:2018ike}.
The suppression arises because dark matter particle velocities tend to be suppressed both close to and far from the subhalo center.
Near the center, there is less enclosed mass and thus the virial velocities of the particles are low; far from the center, the escape velocity is low and sets the ceiling for the velocities of dark matter particles that remain gravitationally bound in the subhalo.

The expected angular size of the gamma-ray emission is given in Table~\ref{tab:J_and_theta}.
We see that $\langle\theta\rangle / \theta_0$ differs from the $s$-wave case ($n=0$) the most if dark matter annihilation is Sommerfeld-enhanced ($n=-1$).
For a single NFW halo with precisely-known profile parameters, distinguishing between these two scenarios may be possible if an instrument could resolve $\langle\theta\rangle$ to within $\sim 30\%$ [distinguishing between models involving $p$-wave ($n=2$) or $d$-wave ($n=4$) annihilation would require more precision].
However, for systems with stars in them, there is always some uncertainty in determining the profile parameters from stellar measurements, even if the potential is dominated by dark matter~\cite{Strigari:2018utn}.
If the uncertainty in $\theta_0$ is $\gtrsim 40\%$, then even a precise measurement of $\langle\theta\rangle$ would not likely be sufficient to distinguish between the $n=-1$ and $n=0$ annihilation models.
In principle, the different choices of the dark matter annihilation model lead to different scale free angular distributions, implying that changing $n$ is not completely degenerate with changing the halo profile parameters $r_s$ and $\rho_s$.
To distinguish between different choices of $n$, observations would need to be able to resolve the full shape of the angular distribution.

\begin{center}
\begin{table}[t]
\begin{tabular}{|c|c|c|c|}
  \hline
  $n$ & Annihilation Model & $\tilde{J}_{S(n)}^\tot$ & $\langle\theta\rangle / \theta_0$ \\\hline
  -1  & Sommerfeld & 1.02 & 0.27 \\\hline
   0  & $s$-wave   & 0.33 & 0.39 \\\hline
   2  & $p$-wave   & 0.15 & 0.45 \\\hline
   4  & $d$-wave   & 0.12 & 0.46 \\\hline
\end{tabular}
\caption{Numerical values for the scale free total effective $J$-factor $\tilde{J}_{S(n)}^\tot$ from Eq.~\eqref{eq:Jtottilde} and the approximate angular size of the gamma-ray emission $\langle\theta\rangle$ in units of $\theta_0 \equiv r_s/D$ from Eq.~\eqref{eq:angular_size}.
  Both quantities depend on the dark matter annihilation model (specified by $n$) and the scale free dark matter velocity distribution function, but they are independent of the NFW profile parameters $r_s$ and $\rho_s$.}
\label{tab:J_and_theta}
\end{table}
\end{center}

\section{Gamma-ray emission from a distribution of subhalos}
\label{sec:subahlo-distribution}

We have thus far focused on the case of gamma-ray emission from dark matter annihilation in a single subhalo.
Let us now apply our methods to the case of emission from an ensemble of Milky Way subhalos.
If we assume all subhalos have the same density profile, then they share a universal scale free dark matter velocity distribution $\tilde{f}(\tilde{r},\tilde{v})$, and all scale free quantities previously introduced are the same for all subhalos.
Thus, changing the annihilation model (i.e., changing $n$) uniformly affects the normalized angular distribution of gamma-ray emission from all subhalos.
The expected angular size of the gamma-ray emission scales as $\langle\theta\rangle / \theta_0$ across all subhalos for different annihilation models.
As seen from Table~\ref{tab:J_and_theta}, using the angular size to distinguish between models is at most a $\sim 40\%$ effect for an NFW profile.

A potentially much more important consideration for model discrimination is comparing the total expected gamma-ray flux from each subhalo.
Let us consider a subhalo distribution $\xi$, obtained from simulation or possibly determined from future gravitational lensing studies~\cite{Hezaveh:2016ltk,Cyr-Racine:2018htu}, that is characterized by the subhalo mass, scale radius $r_s$, and distance away $D$.
Defining the halo scale mass as $M_s \equiv \rho_s r_s^3$, the mass of a halo is given by
\begin{equation}
  M_\halo = M_s \left[4\pi \int_0^{\tilde{r}_\max} d\tilde{r}\ \tilde{r}^2\ \tilde{\rho}(\tilde{r}) \right],
\end{equation}
where $\tilde{r}_\max \equiv r_t/r_s$ is a cut-off set by the tidal radius, $r_t$.
For an NFW profile, $M_\halo/M_s = 4\pi [\ln(1+\tilde{r}_\max) - \tilde{r}_\max/(1+\tilde{r}_\max)]$.
Although there is a mild, logarithmic dependence of $M_\halo$ on $\tilde{r}_\max > 1$, the mass of subhalos within a Milky Way-sized galaxy can vary over a range of up to five orders of magnitude~\cite{Springel:2008cc}, rendering any variation of $\tilde{r}_\max$ relatively insignificant.
We thus neglect the $\tilde{r}_\max$ dependence and use $M_s$ as a proxy for $M_\halo$.
The subhalo distribution is thus $\xi(M_s,r_s,D)$.

We rewrite the total effective $J$-factor for a single halo in Eq.~\eqref{eq:Jtot} as
\begin{equation}
  J_{S(n)}^\tot (M_s,\theta_0,D) = \frac{4\pi M_s^2}{\theta_0^3 D^5}
  \left(\frac{4\pi G_N M_s}{\theta_0 D c^2}\right)^{n/2}
  \tilde{J}_{S(n)}^\tot ,
  \label{eq:Jtot-alt}
\end{equation}
using the angular scale $\theta_0$ and scale mass $M_s$ rather than $r_s$ and $\rho_s$.
The total effective $J$-factor due to all subhalos can then be expressed as
\begin{align}
  \sum_\textrm{subhalo} J_{S(n)}^\tot
  = \int & dM_s\ d\theta_0\ dD\ J_{S(n)}^\tot(M_s,\theta_0,D) \nonumber\\
  & \times D\, \xi(M_s,r_s=\theta_0 D,D),
\end{align}
where the additional factor of $D$ arises from integrating $\xi$ over $\theta_0$ rather than $r_s$.
The quantity $I_n(\theta_0) \equiv \int dM_s~ dD~ J_{S(n)}^\tot (M_s,\theta_0,D)\ D~ \xi(M_s,r_s=\theta_0 D,D)$ is thus a relative measure of the gamma-ray flux arising from all subhalos of nominal size $\theta_0$.

We can now see the effect on the gamma-ray flux arising from distribution of subhalos if we deviate from the case of $s$-wave annihilation ($n=0$).
The flux from all subhalos is rescaled by a factor of $(4\pi G_N/c^2)^{n/2} \tilde{J}_{S(n)}^\tot / \tilde{J}_{S(0)}^\tot$, but rescaling this factor is degenerate with changing $(\sigma v)_0$.
However, the flux from each halo is rescaled by an additional factor of $(M_s / \theta_0 D)^{n/2}$ that varies between halos.
Models of $p$- or $d$-wave annihilation will thus see an enhancement in the flux from subhalos which are closer, more massive, or have a smaller angular size, and a suppression in the flux from subhalos which are farther, less massive, or have a larger angular size.
For the case of Sommerfeld-enhanced annihilation, these considerations are reversed.
The enhancement or suppression of the relative flux is potentially a much more significant effect than the change in angular size of emission in each individual halo, as discussed at the beginning of this section.
For example, the scale radii of subhalos of a Milky Way-sized galaxy can vary by up to two orders of magnitude, and the subhalo masses can vary by up to five orders of magnitude~\cite{Springel:2008cc}.
Thus, the factor $(M_s / \theta_0 D)^{n/2} = (M_s/r_s)^{n/2}$ may be quite large, especially for large $n$.

Thus far, we have made no assumptions regarding any relationship between the dark matter scale density and scale radius.
To establish such a relationship, we appeal to numerical simulations.
Considering larger-mass halos that have not been accreted into a larger mass halo, simulations find the relation $M_{\halo} \propto \sigma^{\alpha}$, where $\sigma$ is the 1D dark matter velocity dispersion and $\alpha \sim 3.3$~\cite{Zahid:2018odi}.
This corresponds to the relationship $M_s \propto r_s^\frac{\alpha}{\alpha -2}$.
Dark matter-only simulations of subhalos that have been accreted into a Milky Way-mass dark matter halo imply a similar relationship, as the mass of the halo is found to scale as the maximum circular velocity~\cite{Springel:2008cc}.
Therefore, although dark matter may be tidally stripped from the outer part of the subhalo, the dark matter velocity distribution remains undisturbed in the interior regions.
Moreover, we note that the subhalo tidal radius is generally much larger than the scale radius.
Since the square of the dark matter density for an NFW profile falls off rapidly outside the scale radius, tidal stripping is largely irrelevant to the gamma-ray emission from dark matter annihilation in a subhalo.
Even if the subhalo mass is suppressed relative to that expected from the relation $M_{\halo} \propto \sigma^{\alpha}$ as a result of tidal stripping, it should not affect the relationship between the subhalo parameters ($M_s \propto r_s^\frac{\alpha}{\alpha -2}$), which in turn determines the subhalo effective $J$-factor.

We use the above relationship between $M_s$ and $r_s$ to express the subhalo velocity distribution as a function with a single parameter.
For dark matter annihilation with a $v^n$ velocity dependence, the emission from a subhalo at a distance $D$ is rescaled from the case of $n=0$ by a factor $\propto (\theta_0 D)^\frac{n}{\alpha-2}$.
In particular, for substructures at a fixed distance, $p$-wave and $d$-wave annihilation increase the relative gamma-ray flux from substructures with a large angular size, while Sommerfeld-enhanced annihilation increases the relative flux from substructures with a small angular size.
This scaling is opposite to that found earlier in this section, without relating $M_s$ and $r_s$.

As an example, we can consider the subhalo distribution function described in Ref.~\cite{Han:2015pua}; under the relevant assumptions, the distribution of halos with scale mass $M_s$ at a distance $r$ from the Galactic Center is given by
\begin{equation}
  \frac{d^2 N}{dM_s dr} \propto M_s^{-2} r^3 \rho_\MW (r),
\end{equation}
where $\rho_\MW (r)$ is the dark matter distribution in the Milky Way halo.
We then have
\begin{align}
  \xi (M_s, r_s, D) \propto & \left[M_s^{-2} \times
  \delta \left(r_s - c_0 M_s^{1-\frac{2}{\alpha}} \right) \right] \nonumber\\
  &\times \left[\frac{1}{2} \int_{-1}^{1} d\cos\alpha\ r^3 \rho_\MW (r) \right],
  \label{eq:HaloDistributionAnsatz}
\end{align}
where $r = (r_{\odot}^2 + D^2 -2r_\odot D \cos \alpha )^{1/2}$, $r_{\odot}$ is the distance from the solar system to the Galactic Center, and $c_0$ is a proportionality constant.
The subhalo distribution function factorizes into a term which depends on position and a term which depends on halo mass.
The $\theta_0$-dependence of $I_n (\theta_0)$ appears only in the mass-dependent term, and for the halo distribution function in Eq.~\eqref{eq:HaloDistributionAnsatz}, is given by
\begin{equation}
  I_n (\theta_0) \propto \theta_0^{-3+\frac{n+2}{\alpha-2}} .
\end{equation}
The contribution to the gamma-ray flux from subhalos decreases with $\theta_0$ for the cases of Sommerfeld-enhanced or $s$-wave annihilation, is nearly independent of $\theta_0$ for the case of $p$-wave annihilation, and grows with $\theta_0$ for the case of $d$-wave annihilation.

\section{Conclusion}
\label{sec:conclusion}

We have considered the effect of velocity-dependent dark matter annihilation on gamma-ray emission from dark matter subhalos.
Our starting assumption is that the dark matter distribution is spherically symmetric with isotropic orbits and depends only on two parameters: the scale radius $r_s$ and scale density $\rho_s$.
We find that the effect of a velocity-dependent dark matter annihilation cross section is largely determined by dimensional analysis, since the only velocity scale in the problem is $(4\pi G_N \rho_s r_s^2)^{1/2}$.
In the case of velocity-dependent dark matter annihilation, the angular size of gamma-ray emission of any individual subhalo is rescaled by a factor which depends on the form of the dark matter distribution, but not on the halo parameters.
But the relative normalization of the gamma-ray flux from different subhalos depends on the parameters of the halo through the combination $(\rho_s r_s^2)^{n/2}$, but not on the form of the dark matter distribution.

We have applied these results to the specific case of an NFW dark matter profile, under the assumption that dark matter annihilation is either $s$-wave, $p$-wave, $d$-wave, or Sommerfeld-enhanced in the Coulomb limit.
We have also considered the application of these results to gamma-ray emission from a distribution of Milky Way subhalos.
In particular, we have found that for substructures at a fixed distance, $p$-wave or $d$-wave annihilation will increase the relative flux from substructures with a larger angular size, while Sommerfeld-enhanced annihilation will increase the relative flux from substructures with a smaller angular size.

Although we have focused on the case of an NFW profile, it is easy to modify these results for other cases involving at most two dimensionful parameters, including the generalized NFW profile, the Einasto profile, etc.
For any such profile, there is a scale free angular distribution $\tilde{J}_{S(n)}(\tilde{\theta})/\tilde{J}_{S(n)}^\tot$, which depends on the velocity dependence of the annihilation cross section and the form of the velocity distribution.
Meanwhile, the dependence on the halo parameters is entirely determined by dimensional analysis.

Extracting the size of the gamma-ray emission from a subhalo is likely challenging for current observatories, such as the Fermi-LAT, though it may be possible for future ones with higher angular resolution, such as the Cerenkov Telescope Array.~\footnote{\url{https://www.cta-observatory.org/}}
In order to extract this signal, it will be crucial to obtain a more precise prediction and/or measurement of the dark matter density profiles in dwarf galaxies and subhalos.
This can be accomplished either via numerical simulations or through measurements of precise radial and tangential velocities of stars.
For example, a measurement of the tangential velocities of stars in dwarf galaxies to a precision of $\sim 1$~km/s with an observatory such as Gaia should be able to distinguish between cored and cusped dark matter density profiles~\cite{Strigari:2018bcn}.

In applying this formalism, we have used the Eddington inversion formula to derive the full velocity distribution from the density profile.
We can thus determine the effect of velocity-dependent (power-law or otherwise) dark matter annihilation on the gamma-ray flux from a subhalo.
Alternatively, we could have used the results of Ref.~\cite{Taylor:2001bq}, which showed using numerical simulations that the dark matter density $\rho$ scales with the 1D velocity dispersion $\sigma$ as $\rho / \sigma^3 \sim r^{-\beta}$, where $\beta \sim 1.875$.
The advantage is that this approach is valid under less restrictive assumptions than the Eddington inversion formula, but the disadvantage is that it only provides the velocity dispersion, not the full velocity distribution, and thus can only be directly applied to the $p$-wave case.
It would be interesting to consider this approach in more detail.

Given an appropriate probability distribution for subhalos in the Milky Way, it would be possible to predict the angular scale of anisotropies in the gamma-ray spectrum arising from dark mater annihilation.
As increasingly refined estimates of these probability distributions are obtained from numerical simulation and gravitational lensing analyses, it would be interesting to compare these predictions to the observed angular distribution of gamma rays.

\vskip 0.1in
{\bf Acknowledgements.}
We are grateful to Francis-Yan Cyr-Racine, Yashar Hezaveh, Manoj Kaplinghat, and Jabran Zahid for useful discussions.
KB acknowledges the support of a Johns Hopkins Provost’s Postdoctoral Fellowship.
The work of JK is supported in part by Department of Energy grant DE-SC0010504.
The work of LES is supported by DOE Grant de-sc0010813.



\end{document}